# Lorentz-breaking Theory and Tunneling Radiation Correction to Vaidya-Banner de Sitter Black Hole

Bei Sha[*] and Zhi-E Liu

Qilu Normal University, No. 2, Wenbo Road, Zhangqiu District, Jinan, China
*. Corresponding author: shabei@qlnu.edu.cn

**Abstract**

In Vaidya-Bonner de Sitter Black hole space-time, the tunneling radiation characteristics of fermions and bosons are corrected by taking Lorentz symmetry breaking theory into account. The corresponding gamma matrices and ether-like field vectors of the black hole are constructed, then the new modified form of Dirac equation for the fermion with spin 1/2 and the new modified form of Klein-Gordon equation for boson in the curved space-time of the black hole are obtained. Through solving the two equations, new and corrected expressions of surface gravity, Hawking temperature and tunneling rate of the black hole are obtained, and the results obtained are also discussed.

## 1. Introduction

In recent years, a series of significant studies have been conducted on quantum tunneling radiation and related contents of various static, stationary and non-stationary black holes [1-15]. These studies have involved the tunneling radiation characteristics of fermions and bosons. Notably, the Lorentz dispersion relation has long been considered fundamental to modern physics, and both the theories of general relativity and quantum field seem to be based on this relation. However, the study on the theory of quantum gravity shows that the Lorentz dispersion relation may need to be modified in the high energy case, which must lead to the correction of the dynamical equations for bosons and fermions. Although a set of theory of dispersion relation in the high energy field has not been effectively established, it is certain that the order of magnitude of this correction term numerical value is Planck scale [16-22]. Considering the role of the Lorentz breaking in curved space-time, necessary corrections have been made to the tunneling radiation of bosons and fermions in static and stationary curved space-times, and some meaningful conclusions have been obtained [23-26]. However, for the case of non-stationary black holes, the problem of tunneling radiation correction to bosons and fermions in whose non-stationary curved space-time has not been investigated deeply, therefore, in this paper the characteristics of fermion's and boson's tunneling radiation from Vaidya-Bonner de Sitter black hole will be studied in detail. By constructing the gamma matrices and the ether-like field vectors in the curved space-time of the black hole, the modified Dirac equation and Klein-Gordon equation in the curved space-time of the black hole will be obtained, and the results obtained through solving the two modified equations will be discussed in depth. The second section below will focus on the construction of gamma matrices of the black hole and the obtaining of Dirac particle's dynamical equation. In the third section, the ether-like vectors associated with the black hole will be constructed and the characteristics of the tunneling radiation of the Dirac particle from the black hole will be studied. In the fourth section, the Klein-Gordon equation will be reconstructed using ether-like vectors according to Lorentz breaking theory, and the properties of the tunneling radiation of the boson from the black hole will be studied. The last section will summarize the conclusions obtained above and will have a further discussion about black hole physics.

## 2. Gamma matrices in Vaidya-Bonner de Sitter black hole space-time and modified Dirac dynamical equation

Fermions with spin 1/2 are Dirac particles. After adding Lorentz symmetry violating term to the action of Dirac particles in flat space-time, the Dirac equation of Lorentz symmetry breaking in flat space-time can be derived by using Hamilton principle [27-29]. To generalize the modified Dirac equation from flat space-time to the Vaidya-Bonner de Sitter curved space-time, we need to determine the corresponding Gamma matrices $\gamma^\mu$ of the space-time of the black hole, and to extend the ordinary derivative to the covariant derivative. Taking into account the effect of Lorentz symmetry violating, the dynamical equation of the Fermion with spin 1/2 in the Vaidya-Bonner de Sitter black hole space-time is

$$\left\{\gamma^\mu D_\mu \left[1+\hbar^2 \frac{a}{m^2}\left(\gamma^\mu D_\mu\right)^2\right]+\frac{b}{\hbar}\gamma^5+c\hbar\left(u^\alpha D_\alpha\right)^2-\frac{m}{\hbar}\right\}\Psi=0, \qquad (2.1)$$

where $u^\mu$ are the ether-like field vectors, $a$, $b$, and $c$ are all small quantities and $\frac{a}{m}$, $\frac{b}{m}$, $\frac{c}{m}$ are constants far less than 1, $\Psi$ is the wave function of the Dirac particle, and $\hbar$ is the Planck constant.

The space-time line element of the Vaidya-Bonner de Sitter black hole is

$$ds^2=-\left(1-\frac{2M}{r}+\frac{Q^2}{r^2}-\chi^2 r^2\right)dv^2+2dvdr+r^2\left(d\theta^2+\sin^2\theta d\phi^2\right), \qquad (2.2)$$

from which the non-zero components of the contravariant metric tensor $g^{\mu\nu}$ are respectively

$$\begin{aligned} g^{rr}&=g^{11}=1-\frac{2M}{r}+\frac{Q^2}{r^2}-\chi^2 r^2 \\ g^{rv}&=g^{10}=g^{01}=1 \\ g^{\theta\theta}&=g^{22}=\frac{1}{r^2} \\ g^{\phi\phi}&=g^{33}=\frac{1}{r^2\sin^2\theta} \end{aligned}. \qquad (2.3)$$

In Eqs. (2.2) and (2.3), $M=M(v)$, $Q=Q(v)$, where $v$ is the advanced Eddington coordinate, $\chi$

is the parameter related to de Sitter space. According to Eqs. (2.2) and (2.3), the constructed gamma matrices $\gamma^\mu$ in Eq. (2.1) are

$$\gamma^\nu = \frac{1}{\sqrt{g^{rr}}}\left[i\begin{pmatrix} I & 0 \\ 0 & -I \end{pmatrix} + \begin{pmatrix} 0 & \sigma^1 \\ \sigma^1 & 0 \end{pmatrix}\right] = \frac{1}{\sqrt{g^{rr}}}\begin{pmatrix} i & 0 & 0 & 1 \\ 0 & i & 1 & 0 \\ 0 & 1 & -i & 0 \\ 1 & 0 & 0 & -i \end{pmatrix}$$

$$\gamma^r = \sqrt{g^{rr}}\begin{pmatrix} 0 & \sigma^1 \\ \sigma^1 & 0 \end{pmatrix} = \sqrt{g^{rr}}\begin{pmatrix} 0 & 0 & 0 & 1 \\ 0 & 0 & 1 & 0 \\ 0 & 1 & 0 & 0 \\ 1 & 0 & 0 & 0 \end{pmatrix}$$

$$\gamma^\theta = \frac{1}{\sqrt{g^{\theta\theta}}}\begin{pmatrix} 0 & \sigma^2 \\ \sigma^2 & 0 \end{pmatrix} = \frac{1}{r}\begin{pmatrix} 0 & 0 & 0 & -i \\ 0 & 0 & i & 0 \\ 0 & -i & 0 & 0 \\ i & 0 & 0 & 0 \end{pmatrix}$$

$$\gamma^\phi = \frac{1}{\sqrt{g^{\phi\phi}}}\begin{pmatrix} 0 & \sigma^3 \\ \sigma^3 & 0 \end{pmatrix} = \frac{1}{r\sin\theta}\begin{pmatrix} 0 & 0 & 1 & 0 \\ 0 & 0 & 0 & -1 \\ 1 & 0 & 0 & 0 \\ 0 & -1 & 0 & 0 \end{pmatrix}, \quad (2.4)$$

where $I$ is a unit matrix, and $\sigma^1$, $\sigma^2$, $\sigma^3$ are Pauli matrices. Obviously, Eqs. (2.4) satisfy the following relations

$$\gamma^\mu\gamma^\nu + \gamma^\nu\gamma^\mu = \{\gamma^\mu,\gamma^\nu\} = 2g^{\mu\nu}I. \quad (2.5)$$

In Eq. (2.1), the following relationship exists

$$D_\mu = \partial_\mu + \frac{i}{\hbar}qA_\mu + \frac{i}{2}\Gamma_\mu^{\alpha\beta}\Pi_{\alpha\beta}$$
$$\Pi_{\alpha\beta} = \frac{i}{4}\left[\gamma^\alpha,\gamma^\beta\right] \quad (2.6)$$

where, $\Gamma_\mu^{\alpha\beta}$ is the spin connection, $\frac{i}{2}\Gamma_\mu^{\alpha\beta}\Pi_{\alpha\beta}$ is the rotational contact term, which can be ignored in semi-classical theory. In Eq. (2.6), $\gamma^5$ are the Lorentz spinor generators. In Eq. (2.1), there is the following relationship between $\gamma^\mu$ or $\gamma^\nu$ and $\gamma^5$

$$\gamma^5\gamma^\mu + \gamma^\mu\gamma^5 = 0. \quad (2.7)$$

According to Eqs. (2.4) and (2.7), the expression of $\gamma^5$ is constructed as following

$$\gamma^5 = r^2 \sin\theta \left(\gamma^v \gamma^r - \gamma^r \gamma^v\right) \gamma^\theta \gamma^\phi. \tag{2.8}$$

According to semi-classical theory, for fermion with spin 1/2, the wave function in Eq. (2.1) can be expressed as

$$\Psi = \begin{pmatrix} \xi \\ \eta \end{pmatrix} \exp\left(\frac{i}{\hbar} S\right). \tag{2.9}$$

Substituting Eq. (2.9) into Eq. (2.1), we get

$$i\gamma^\mu \left(\partial_\mu S + qA_\mu\right) \left[1 - \frac{a}{m^2} \gamma^\mu \gamma^\nu \left(\partial_\mu S + qA_\mu\right)\left(\partial_\nu S + qA_\nu\right)\right] \Psi$$
$$+ \left[b\gamma^5 - cu^\mu u^\nu \left(\partial_\mu S + qA_\mu\right)\left(\partial_\nu S + qA_\nu\right) - m\right]\Psi = 0 \tag{2.10}$$

For this black hole, the four-dimensional electromagnetic potential vector $A_\mu = (A_v, 0, 0, 0)$, and

$$A_v = A_0 = \frac{Q}{r}. \tag{2.11}$$

Multiply both sides of Eq. (2.10) by $i\gamma^\nu \left(\partial_\nu S + qA_\nu\right)\left[1 - \frac{a}{m^2} \gamma^\nu \gamma^\mu \left(\partial_\nu S + qA_\nu\right)\left(\partial_\mu S + qA_\mu\right)\right]$ and ignore small amount b/m and use Eq. (2.7) to obtain

$$g^{\mu\nu}\left(\partial_\mu S + qA_\mu\right)\left(\partial_\nu S + qA_\nu\right)(1 + 2a) + 2cmu^\mu u^\nu \left(\partial_\mu S + qA_\mu\right)\left(\partial_\nu S + qA_\nu\right) + m^2 = 0, \tag{2.12}$$

which is the dynamical equation for a Dirac particle with spin 1/2, mass $m$ and charge $q$, and which is derived from Eq. (2.1) by taking into account the prerequisites of Lorentz symmetry violating. Eq. (2.12) and Eq. (2.1) are two equivalent equations. We only need to find the particle's action $S$ from Eq. (2.12), then according to the semi-classical theory and the WKB approximation theory, we can work out the new and corrected quantum tunneling rate and other important physical quantities that describe the black hole.

## 3. Correction to the characteristics of fermion's tunneling radiation from Vaidya-Bonner de Sitter black hole

In order to solve Eq. (2.12), the correct ether-like field vectors $u^\mu$ must be constructed according to Eqs. (2.2), (2.3) and (2.4). As is known to all, in the flat space-time of the canonical coordinate system, the ether-like field vectors $u^\mu$ are constant vectors and satisfy the condition of $u^\mu u_\mu = \text{const}$, but in the curved space-time of the black hole, $u^\mu$ are not constant vectors, but nevertheless, we still request

$$u^\mu u_\mu = \text{const}, \tag{3.1}$$

only according to which and Eqs. (2.2) and (2.3), the constructed $u^\mu$ or $u^\nu$ are correct. Based on Eqs. (2.2), (2.3) and (3.1), the ether-like field vectors can be constructed as

$$u^\nu = u^0 = \frac{c_v}{\sqrt{g_{00}}} = \frac{c_v}{\sqrt{-\left(1 - \frac{2M}{r} + \frac{Q^2}{r} - \chi^2 r^2\right)}}$$

$$u^r = u^1 = \frac{c_r \sqrt{g_{00}}}{\sqrt{g_{01}}} = \frac{c_r}{\sqrt{g_{01}}} \sqrt{-\left(1 - \frac{2M}{r} + \frac{Q^2}{r} - \chi^2 r^2\right)}. \tag{3.2}$$

$$u^\theta = u^2 = \frac{c_\theta}{\sqrt{g_{22}}} = \frac{c_\theta}{r}$$

$$u^\phi = u^3 = \frac{c_\phi}{\sqrt{g_{33}}} == \frac{c_\phi}{r \sin\theta}$$

Obviously, Eq. (3.2) completely satisfy Eq. (3.1). In Eq. (3.2), $c_v$, $c_r$, $c_\theta$ and $c_\phi$ are real constants. Substituting Eqs. (2.3) and (3.2) into Eq. (2.12), we can obtain

$$(1+2a)\left[g^{11}(\partial_r S)^2 + 2g^{10}(\partial_r S)(\partial_v S + qA_0)\right]$$

$$+2cm\left[\frac{c_v^2}{g_{00}}(\partial_v S + qA_0)^2 + \frac{g_{00} c_r^2}{g_{01}}(\partial_r S)^2 + 2\frac{c_v c_r}{\sqrt{g_{01}}}(\partial_v S + qA_0)(\partial_r S)\right] + s_0 + m^2 = 0, \tag{3.3}$$

where $s_0$ is the constant related to $\theta$ and $\phi$ that appears in the process of separating variables.

To solve Eq. (3.3), the general tortoise coordinate transformation must be done, i.e.

$$r_* = r + \frac{1}{2\kappa}\ln\frac{r - r_H(v)}{r_h(v_0)}$$

$$v_* = v - v_0 \tag{3.4}$$

From this transformation we can get

$$\frac{\partial}{\partial r} = \frac{1 + 2\kappa(r - r_H)}{2\kappa(r - r_H)}\frac{\partial}{\partial r_*}$$

$$\frac{\partial}{\partial v} = \frac{\partial}{\partial v_*} - \frac{\dot{r}_H}{2\kappa(r - r_H)}\frac{\partial}{\partial r_*}, \tag{3.5}$$

in which, $\dot{r}_H = \frac{dr_H(v)}{dv}$. Since this black hole is spherically symmetric, the variables of the particle's action $S$ can be separated as

$$S = R(v_*, r_*) + Y(\theta, \phi). \tag{3.6}$$

Let

$$\frac{\partial R}{\partial v_*} = \frac{\partial S}{\partial v_*} = -\omega, \tag{3.7}$$

where $\omega$ is the energy of the particle. Now substitute Eqs. (3.4), (3.5), (3.6) and (3.7) into Eq. (3.3), and at $r \to r_H$, $v \to v_0$ to simplify the obtained equation preliminarily, we get

$$\left[(1+2a)\left(g^{11}-2g^{10}\dot{r}_H\right)+2cm\left(\frac{c_v^2}{g_{00}}\dot{r}_H^2+\frac{g_{00}c_r^2}{g_{01}}-2\frac{c_vc_r}{\sqrt{g_{01}}}\right)\right]\frac{1}{2\kappa(r-r_H)}\left(\frac{\partial R}{\partial r_*}\right)^2$$
$$+2\left[(1+2a)g^{10}-2cm\left(\frac{c_v^2}{g_{00}}\dot{r}_H-\frac{c_vc_r}{\sqrt{g_{01}}}\right)\right](-\omega+qA_0)\left(\frac{\partial R}{\partial r_*}\right)=0 \quad (3.8)$$

Now consider the zero hypersurface equation of the black hole, i.e.

$$g^{\mu\nu}\frac{\partial F}{\partial x^\mu}\frac{\partial F}{\partial x^\nu}=0. \quad (3.9)$$

Substituting Eq. (2.3) into Eq. (3.9), the equation of the horizon of the Vaidya-Bonner de Sitter black hole is obtained

$$1-\frac{2M}{r}+\frac{Q^2}{r^2}-\chi^2r^2-2\dot{r}=0. \quad (3.10)$$

At the event horizon, Eq. (3.10) can be expressed as

$$r_H^2-2Mr_H+Q^2-\chi^2r_H^4-2\dot{r}_Hr_H^2=0. \quad (3.11)$$

For the cosmic horizon, we have

$$r_c^2-2Mr_c+Q^2-\chi^2r_c^4-2\dot{r}_cr_c^2=0. \quad (3.12)$$

To solve Eq. (3.8), the horizons of the black hole must be considered, only in which can we study the characteristics of fermion tunneling radiation at the horizons of the black hole and its related physical significance. Therefore, continue considering the case of $r \to r_H$, $v \to v_0$, and simplifying the form of the Eq. (3.8), we can get

$$\lim_{\substack{r \to r_H \\ v \to v_0}}\frac{A}{B}\left(\frac{\partial R}{\partial r_*}\right)^2-2(\omega-\omega_0)\frac{\partial R}{\partial r_*}=0, \quad (3.13)$$

where

$$\omega_0 = qA_0, \quad (3.14)$$

$$\lim_{\substack{r \to r_H \\ v \to v_0}}\frac{A}{B}=\lim_{\substack{r \to r_H \\ v \to v_0}}\frac{(1+2a)\left(g^{11}-2g^{10}\dot{r}_H\right)+2cm\left(\frac{g_{00}c_r^2}{g_{01}}-2\frac{c_vc_r}{\sqrt{g_{01}}}\dot{r}_H+\frac{c_v^2}{g_{00}}\dot{r}_H^2\right)}{2\kappa(r-r_H)\left[(1+2a)g^{10}-2cm\frac{c_v^2}{g_{00}}\dot{r}_H+2cm\frac{c_vc_r}{\sqrt{g_{01}}}\right]}. \quad (3.15)$$

Substituting the expressions of $g_{00}$ and $g_{01}$ in Eq. (2.2) and the expression of $g^{11}$ in Eq. (2.3) into Eq. (3.15), we get

$$\lim_{\substack{r\to r_H \\ v\to v_0}} \frac{A}{B} = \lim_{\substack{r\to r_H \\ v\to v_0}} \frac{(1+2a)\left(1-2Mr^{-1}+Q^2r^{-2}-\chi^2r^2-2\dot{r}_H\right)-cm\left[c_v^2+4c_r^2+4c_vc_r\right]\dot{r}_H}{2\kappa[r-r_H(v)]\left(1+2a+cmc_v^2+2cmc_vc_r\right)}. \quad (3.16)$$

Ignoring small quantities of second order $\left[c_v^2+4c_r^2+4c_vc_r\right]$ in Eq. (3.16), then when $r\to r_H$, $v\to v_0$, both the numerator and the denominator of Eq. (3.16) approach 0, let

$$\lim_{\substack{r\to r_H \\ v\to v_0}} \frac{A}{B} = \lim_{\substack{r\to r_H \\ v\to v_0}} \frac{(1+2a)\left(1-2Mr^{-1}+Q^2r^{-2}-\chi^2r^2-2\dot{r}_H\right)}{2\kappa[r-r_H(v)]\left(1+2a+cmc_v^2+2cmc_vc_r\right)} = 1. \quad (3.17)$$

Using L'Hopital's rule, then we get

$$\kappa = \left(\frac{M}{r_H^2}-\frac{Q^2}{r_H^3}-\chi^2 r_H\right)\left[1-\tilde{m}+\tilde{m}^2-\cdots\right], \quad (3.18)$$

where

$$\tilde{m} = \frac{c\left(c_v^2+2c_vc_r\right)}{1+2a}m. \quad (3.19)$$

It can be seen from Eqs. (3.18) and (3.19) that the surface gravity $\kappa$ is related to $c$, $a$, $c_v$, and $c_r$, and that the introduced ether-like field vectors and the correction terms of gamma matrices considering Lorentz breaking are equivalent to the change of the mass of the particle, which also demonstrates the correlation of mass and space-time. Substituting Eq. (3.17) into Eq. (3.13), we get

$$\frac{\partial R_\pm}{\partial r_*} = (\omega-\omega_0)\pm(\omega-\omega_0). \quad (3.20)$$

According to Eq. (3.5), it can be obtained that

$$\frac{\partial R_\pm}{\partial r} = \frac{1+2\kappa(r-r_H)}{2\kappa(r-r_H)}\frac{\partial R_\pm}{\partial r_*} = \frac{1+2\kappa(r-r_H)}{2\kappa(r-r_H)}\left[(\omega-\omega_0)\pm(\omega-\omega_0)\right], \quad (3.21)$$

using residue theorem, which can be solved that

$$R_\pm = \frac{i\pi}{2\kappa}\left[(\omega-\omega_0)\pm(\omega-\omega_0)\right]. \quad (3.22)$$

According to the quantum tunneling radiation theory and the semi-classical theory, the tunneling rate of fermion with spin 1/2 at the event horizon of the Vaidya-Bonner de Sitter black hole can be expressed as

$$\Gamma \sim \exp(-2\operatorname{Im} S_\pm) = \exp(-2\operatorname{Im} R_\pm)$$
$$= \exp\left[-\frac{2\pi}{\kappa}(\omega-\omega_0)\right] = \exp\left(-\frac{\omega-\omega_0}{T_H}\right), \quad (3.23)$$

where

$$T_H = \frac{\kappa}{2\pi} = \frac{1}{2\pi}\left(\frac{M}{r_H^2} - \frac{Q^2}{r_H^3} - \chi^2 r_H\right)\left[1 - \tilde{m} + \tilde{m}^2 - \cdots\right], \tag{3.24}$$
$$= T_0\left[1 - \tilde{m} + \tilde{m}^2 - \cdots\right]$$

where

$$T_0 = \frac{1}{2\pi}\left(\frac{M}{r_H^2} - \frac{Q^2}{r_H^3} - \chi^2 r_H\right). \tag{3.25}$$

Here $T_0$ is the Hawking temperature at the event horizon of the black hole before correction, while $T_H$ is the Hawking temperature at the event horizon of the black hole after correction, which is obtained by considering Lorentz symmetry violation and ether-like field vectors introduced. It can be seen that Lorentz-breaking Dirac field affects the Hawking temperature, the surface gravity and the tunneling rate of the Dirac particle at the black hole event horizon. Eqs. (3.18), (3.23) and (3.24) are new and corrected expressions.

This black hole also has a cosmic horizon $r_c$. Using the same research method above, we can get the surface gravity

$$\kappa_c = \left(\frac{M}{r_c^2} - \frac{Q^2}{r_c^3} - \chi^2 r_c\right)\left[1 - \tilde{m} + \tilde{m}^2 - \cdots\right], \tag{3.26}$$

the tunneling rate of the spin 1/2 fermion at the cosmic horizon of this black hole,

$$\Gamma^c \sim \exp\left(-\frac{\omega - \omega_0^c}{T_c}\right), \tag{3.27}$$

where $T_c$ is the Hawking temperature at the cosmic event horizon of the black hole,

$$T_c = \frac{\kappa_c}{2\pi} = T_0^c\left[1 - \tilde{m} + \tilde{m}^2 - \cdots\right], \tag{3.28}$$

where

$$T_0^c = \frac{1}{2\pi}\left(\frac{M}{r_c^2} - \frac{Q^2}{r_c^3} - \chi^2 r_c\right). \tag{3.29}$$

$\omega_0^c$ in Eq. (3.27) is

$$\omega_0^c = \frac{eQ}{r_c}. \tag{3.30}$$

It can be seen that the coefficient $a$ corresponding to gamma matrices $\gamma^\mu$ and the coefficient $c$ corresponding to the ether-like field vectors term both have influences on the tunneling rate and Hawking temperature at the event and cosmic horizons of the black hole, so, which is an important topic that must be paid attention to in the current research on the thermodynamic evolution of black holes.

For the fermion with spin 1/2 in the space-time of Vaidya-Banner de Sitter black hole, we can add the coupling term of Lorentz symmetry violating into the action of Dirac particle, based on which to study the characteristics of the Dirac particle's radiation from this black hole, while for fermions with spin 3/2 and so on, we can study them in the same way, and correct the physical quantities such as the tunneling radiation rate and Hawking temperature etc., starting from the Rarita-Schwinger equation.

## 4. Correction to the characteristics of boson's tunneling radiation from Vaidya-Bonner de Sitter black hole

For the bosons, we can rewrite the action of the scalar field in this black hole space-time, still considering the coupling term of Lorentz symmetry violating. The action of the boson with mass $m$ is [30-31]

$$L^B = \int d^4x \sqrt{-g} \frac{1}{2}\left[ \Phi \Box \Phi + \lambda \left(u^\mu \partial_\mu \Phi\right)^2 + m^2 \Phi^2 \right], \tag{4.1}$$

where $\lambda$ is the coefficient of the correction term, and is a small constant. $u^\mu$ are ether-like field vectors, just as shown in Eq. (3.2). Here, it still must be required to satisfy Eq. (3.1). In Eq. (4.1)

$$\begin{aligned}\Box \Phi &= \nabla \cdot (\nabla \Phi) = \nabla_\mu \left( g^{\mu\nu} \frac{\partial \Phi}{\partial x^\nu} \right) \\ &= \frac{1}{\sqrt{-g}} \frac{\partial}{\partial x^\mu} \left( \sqrt{-g}\, g^{\mu\nu} \frac{\partial \Phi}{\partial x^\nu} \right)\end{aligned}, \tag{4.2}$$

$$\nabla_\mu B^\mu = B^\mu_{;\mu} = B^\mu_{,\mu} + \Gamma^\mu_{\alpha\mu} B^\alpha = \frac{1}{\sqrt{-g}} \frac{\partial}{\partial x^\mu}\left(\sqrt{-g}\, B^\mu\right), \tag{4.3}$$

$$\Gamma^\mu_{\alpha\mu} = \frac{1}{2} g^{\mu\nu} g_{\mu\nu,\alpha} = \frac{\partial}{\partial x^\alpha}\left(\ln\sqrt{-g}\right). \tag{4.4}$$

The dynamical equation of bosons is determined by the following equation, i.e.

$$\delta L^B = 0. \tag{4.5}$$

Therefore, the dynamical equation of the boson with mass $m$ in the space-time of this black hole is

$$\frac{1}{\sqrt{-g}} \frac{\partial}{\partial x^\mu}\left[\sqrt{-g}\left(g^{\mu\nu} + \lambda u^\mu u^\nu\right)\frac{\partial}{\partial x^\nu}\right]\Phi + m^2 \Phi = 0. \tag{4.6}$$

This equation is actually the modified Klein-Gordon equation that takes the Lorentz symmetry breaking into account. By the same theoretical method, for the boson with mass $m$ and charge $e$, the dynamical equation in the space-time of the black hole is

$$\frac{1}{\sqrt{-g}}\left(\frac{\partial}{\partial x^\mu} - ieA_\mu\right)\left[\sqrt{-g}\left(g^{\mu\nu} + \lambda u^\mu u^\nu\right)\left(\frac{\partial}{\partial x^\nu} - ieA_\nu\right)\right]\Phi + m^2 \Phi = 0. \tag{4.7}$$

Here $\Phi$ is the boson's wave function, $A_\mu$ is the electromagnetic potential vectors. The relation between $\Phi$ and particle's action $S$ is expressed as

$$\Phi = \Phi_0 \exp\left(\frac{i}{\hbar} S\right). \tag{4.8}$$

Substituting Eq. (4.8) into Eq. (4.7), the dynamical equation of the boson with charge $e$ and mass $m$ is obtained

$$\left(g^{\mu\nu} + \lambda u^\mu u^\nu\right)\left(\frac{\partial S}{\partial x^\mu} - eA_\mu\right)\left(\frac{\partial S}{\partial x^\nu} - eA_\nu\right) - m^2 = 0. \tag{4.9}$$

Substituting Eqs. (2.3), (2.11) and (3.2) into Eq. (4.9) and separating variables, the dynamical equation of the boson with mass $m$ and charge $e$ in the Vaidya-Banner de Sitter curved space-time is simplified as

$$\begin{aligned} & g^{11}\left(\frac{\partial S}{\partial r}\right)^2 + 2g^{01}\left(\frac{\partial S}{\partial r}\right)\left(\frac{\partial S}{\partial v} - eA_0\right) \\ & + \lambda\left[\frac{c_v^2}{g_{00}}\left(\frac{\partial S}{\partial v} - eA_0\right)^2 + \frac{c_r^2 g_{00}}{g_{01}}\left(\frac{\partial S}{\partial r}\right)^2 + 2\frac{c_v c_r}{\sqrt{g_{01}}}\left(\frac{\partial S}{\partial r}\right)\left(\frac{\partial S}{\partial v} - eA_0\right)\right] + \eta_0 - m^2 = 0 \end{aligned} \tag{4.10}$$

where $\eta_0$ is the constant introduced in the process of separating variables. Next we solve the Eq. (4.10) applying the general tortoise coordinate transformation just like we did above, we can get the boson's surface gravity

$$\kappa^B = \frac{1}{1 + \frac{1}{2}\lambda c_v^2 \dot{r}_H + \lambda c_v c_r}\left(\frac{M}{r_H^2} - \frac{Q^2}{r_H^3} - \chi^2 r_H\right), \tag{4.11}$$

the tunneling rate at the event horizon of the black hole

$$\Gamma^B \sim \exp(-2\,\text{Im}\,S_\pm) = \exp\left[-\frac{2\pi}{\kappa^B}(\omega - \omega_0)\right] = \exp\left(-\frac{\omega - \omega_0}{T_H^B}\right), \tag{4.12}$$

where

$$\omega = -\frac{\partial R}{\partial v_*} = -\frac{\partial S}{\partial v_*}, \tag{4.13}$$

$$\omega_0 = -eA_0, \tag{4.14}$$

and the Hawking temperature

$$T_H^B = \frac{\kappa^B}{2\pi} = \frac{1}{2\pi\left[1 + \frac{1}{2}\lambda c_v^2 \dot{r}_H + \lambda c_v c_r\right]}\left(\frac{M}{r_H^2} - \frac{Q^2}{r_H^3} - \chi^2 r_H\right). \tag{4.15}$$

Obviously, considering the influence of Lorentz symmetry breaking, the boson's tunneling rate and Hawking temperature at the event horizon of the black hole must be properly corrected, only that the Hawking temperature and other physical parameters obtained can be correct. It should be noted that the physical quantities such as the tunneling rate of bosons and the Hawking temperature at the cosmic horizon $r_c$ of the black hole should also be properly corrected. Using the same method above, we can obtain the boson's tunneling rate and the Hawking temperature at the cosmic horizon of the black hole. The results are simply to replace the $r_H$ in Eqs. (4.12) and (4.15) with $r_c$.

## 5. Conclusions and discussions

Lorentz symmetry breaking is an important subject worthy of further study, searching for which is one of the most sensitive ways of looking for new physics, either new interactions or modifications of known ones [32]. In this paper, we studied and corrected the quantum tunneling radiation properties of fermion and boson at the horizons of Vaidya-Bonner de Sitter black hole based on the Lorentz breaking theory. First, we constructed the gamma matrices [Eqs. (2.4) and (2.7)] and the ether-like field vectors [Eq. (3.2)] of the black hole, which is the key point to study this topic in our method, and then the new and modified form of Dirac equation was obtained. By solving the modified Dirac equation, the surface gravity [Eqs. (3.18) and (3.26)], the Hawking temperature [Eqs. (3.24) and (3.28)] and the tunneling rate [Eqs. (3.23) and (3.27)] of the fermion with spin 1/2 at the event horizon and the cosmic horizon were obtained. For the fermion with spin 3/2 and so on, it can be studied in the same way only need starting from the Rarita-Schwinger equation. For the boson, we got the new form of the Klein-Gordon equation by the constructed ether-like field vectors [Eq. (3.2)], by solving which we obtained the tunneling radiation properties [Eqs. (4.11), (4.12) and (4.15)] of the boson at the event horizon and the cosmic horizon of the black hole. Eqs. (3.18), (3.24), (3.23) and Eqs. (4.11), (4.12) and (4.15) are new and corrected expressions of the tunneling radiation properties of this black hole. We can see that the tunneling radiation properties of fermions are different from those of bosons. Meanwhile, if the correction terms in the expressions obtained above about the characteristics of the tunneling radiation of the black hole are get rid of, the results will be completely consistent with the original results, which also proves that the obtained results of the tunneling radiation characteristics of fermions and bosons in curved space-time of the Vaidya-Bonner de Sitter black hole are correct.

Another important physical concept of black hole physics is entropy $S_{BH}$ of the black hole. From the expressions above about the tunneling rate of fermions and bosons and the Hawking temperature at the black hole horizons, we can know that Lorentz symmetry breaking has a certain effect on them, which will inevitably lead to the change of the black hole entropy. Here $\Delta S_{BH}$ is employed to express the change of Bekenstein-Hawking entropy $S_{BH}$, so the tunneling rate can be expressed as $\Gamma \sim e^{\Delta S_{BH}}$. Black hole entropy is also a kind of subject worthy of study, which can help us to understand the thermodynamic evolution of black holes and related problems deeply.

## Acknowledgments

During the research of this topic, Professor Shu-Zheng Yang had a very useful discussion with us, and we would like to express our great thanks.

This work is supported by the National Nature Science Foundation of China (No. 11273020 and No. U2031121) and Shandong Provincial Nature Science Foundation, China (No. ZR2019MA059).